\documentclass[
 reprint,
 amsmath,amssymb,
 aps,
]{revtex4-2}

\usepackage{graphicx}
\usepackage{dcolumn}
\usepackage{bm}
\usepackage{booktabs}
\usepackage[colorlinks=true,allcolors=blue]{hyperref}

\DeclareRobustCommand{\VAN}[3]{#2}
\let\VANthebibliography\thebibliography
\def\thebibliography{\DeclareRobustCommand{\VAN}[3]{##3}\VANthebibliography}

\begin{document}

\title{Probing Dynamical Dark Energy with Hyper Suprime-Cam Year 3 Cosmic Shear}

\author{Shreya Mukherjee}
 \email{shreya.mukherjee@iucaa.in}
 \affiliation{The Inter-University Centre for Astronomy and Astrophysics, Post bag 4, Ganeshkhind, Pune 411007, India}
\author{Surhud More}
 \email{surhud@iucaa.in}
\affiliation{The Inter-University Centre for Astronomy and Astrophysics, Post bag 4, Ganeshkhind, Pune 411007, India}
\affiliation{Kavli Institute for the Physics and Mathematics of the Universe (WPI), The University of Tokyo Institutes for Advanced Study (UTIAS), The University of Tokyo, Chiba 277-8583, Japan}

\date{\today}

\begin{abstract}
Recent measurements of baryon acoustic oscillations (BAO) and Type Ia Supernovae (SNe) have revealed a preference for a time-evolving dark energy equation of state. Because such dynamical behavior alters both the cosmic expansion history and the late-time growth of structure, weak gravitational lensing measurements can be used to test these extended cosmological models. In this work, we present an extended cosmological analysis of the real-space cosmic shear two-point correlation functions (2PCFs) from the Hyper Suprime-Cam Year 3 (HSC-Y3) survey. Utilizing a shape catalog comprising over 25 million galaxies across 416 deg$^2$, divided into four tomographic redshift bins ($0.3 < z \le 1.5$), we constrain both a constant ($w$CDM) and a time-varying ($w_0w_a$CDM) dark energy equation of state. To break the geometric degeneracies inherent in cosmic shear, we combine the HSC-Y3 data with a CMB-independent suite of background distance probes: DESI DR2 BAO, the recalibrated DES five-year SN sample (DOVEKIE), and Big Bang Nucleosynthesis (BBN) priors. In the $w$CDM paradigm, our joint analysis yields $w =-0.936^{+0.074}_{-0.067}$, remaining statistically consistent with a cosmological constant. Within the $w_0w_a$CDM framework, we find $w_0 =-0.832^{+0.071}_{-0.059}$ and $w_a = -1.23^{+0.66}_{-0.44}$, representing a $2.4\sigma$ preference for dynamical dark energy. Although the unconstrained dark energy parameters reduce the cosmic shear-only $S_8$ constraints to smaller values, our joint multi-probe analysis agrees with the baseline HSC-Y3 $\Lambda$CDM structure growth amplitude. We obtain stable joint constraints of $S_8 = 0.774^{+0.027}_{-0.034}$ for $w$CDM and $S_8 = 0.772^{+0.031}_{-0.030}$ for $w_0w_a$CDM. Our results demonstrate that the weak lensing amplitude is robust to extended dark energy parameterizations, and that the dynamical dark energy signal is currently driven primarily by the geometric distance ladders, a critical consistency test for next-generation Stage-IV cosmological surveys.
\end{abstract}
                           
\maketitle

\section{Introduction}

The flat $\Lambda$ Cold Dark Matter ($\Lambda$CDM) model has served as the concordance model of cosmology for over two decades. It successfully explains a remarkably diverse set of astrophysical observations, including the fluctuations in the Cosmic Microwave Background (CMB) \citep{2020A&A...641A...6P}, the abundances of light elements from Big Bang Nucleosynthesis (BBN) \citep{Fields_2020}, and the large-scale structure of the late-time Universe probed by galaxy clustering and weak gravitational lensing \citep{Hinshaw_2013, 2020, Hamana_2020, Asgari_2021}. However, as observational precision has transitioned into an unprecedented regime, intriguing discrepancies have started to emerge between the early-universe predictions of the flat $\Lambda$CDM model and late-universe measurements. One of the most persistent challenges is the so-called $S_8$ tension. Cosmological models inferred from late-time large-scale structure probes frequently exhibit a lower value for the clustering amplitude parameter, $S_8 \equiv \sigma_8 \sqrt{\Omega_{\rm m}/0.3}$, than what is extrapolated from the Planck CMB measurements \citep{Betoule_2014, Secco_2022, Pantos_2026, Heymans_2021, Abbott_2022}.

While a statistically significant $S_8$ discrepancy could point toward unmodeled baryonic feedback or unknown systematics in the data \citep{Sugiyama_2022, 2022PhRvD.106h3520M}, it also serves as a strong motivation to explore physics beyond the standard model. Recent advancements in the literature have further fueled this pursuit. While $\Lambda$CDM assumes dark energy is a cosmological constant with an equation of state $w = -1$, observations of geometric nature have begun to test this assumption further in a stringent manner. Notably, the Dark Energy Spectroscopic Instrument (DESI) collaboration's Year 1 analysis of Baryon Acoustic Oscillations (BAO), when combined with CMB and Type Ia supernovae data, has shown a mild but compelling preference for a dynamical dark energy equation of state \citep{2025PhRvD.112h3515A}. Parameterized by the Chevallier-Polarski-Linder (CPL) model, $w(a) = w_0 + (1 - a)w_a$ \citep{CHEVALLIER_2001, Linder_2003},  these recent findings suggest that the nature of dark energy may be more complex than a simple cosmological constant. Furthermore, recent re-evaluations and joint analyses from late-time surveys, such as the Kilo-Degree Survey (KiDS), have shown that variations in modeling choices and extended parameter spaces can significantly impact the severity of the $S_8$ tension, bringing low-redshift probes into closer alignment with Planck \citep{Abdalla_2022, 2025A&A...703A.158W, 2026arXiv260114559D, 2026MNRAS.548ag632P}. 

Weak gravitational lensing is exceptionally well-positioned to constrain these extended cosmological models. As the coherent distortion of background galaxy images by foreground matter, cosmic shear is sensitive to both the geometry of the Universe and the growth rate of cosmic structure \citep{BARTELMANN2001291}. Therefore, it provides a direct test of both the dark energy equation of state ($w_0, w_a$) and the clustering amplitude ($S_8$). The Hyper Suprime-Cam (HSC) survey, an optical imaging survey utilizing the 8.2-meter Subaru telescope, is one of the premier Stage-III weak lensing experiments \citep{Aihara_2017, Aihara_2022, Kilbinger_2015}. Its unique combination of depth and excellent seeing allows for precise cosmic shear measurements up to $z \sim 2$. The recent HSC Year 3 (HSC-Y3) real-space cosmic shear analysis \citep{2023PhRvD.108l3518L} provided robust constraints on $S_8$ and $\Omega_{\rm m}$ under the assumption of a flat $\Lambda$CDM cosmology, utilizing a meticulously tested shear catalog \citep{Li_2022} and rigorous mitigation of systematic uncertainties. However, the baseline HSC-Y3 analysis did not explore the parameter space of dynamical dark energy \citep{Li_2022, Zhang_2023, Rau_2023, 2023PhRvD.108l3518L, Dalal_2023, 2023PhRvD.108l3520M, Sugiyama_2023, 2022PhRvD.106h3520M, Chen_2025, Chiu_2024, 2024PhRvD.110h3511S, 2026PhRvD.113j3529Z, 2026PhRvD.113j3530Z, 2025PhRvD.111f3509T, Rana_2026}. 

In this paper, we extend the real-space cosmic shear analysis of the HSC-Y3 dataset to test the $w_0w_a$CDM cosmological model. While the original HSC-Y3 $\Lambda$CDM analysis was conducted under strict blinding protocols to prevent confirmation bias, the extension presented in this work is an unblinded analysis, utilizing the validated HSC-Y3 two-point correlation functions (2PCFs) and covariance matrices. We model the 2PCFs incorporating twenty-three cosmological, astrophysical, and nuisance parameters, utilizing nested Bayesian sampling to constrain the time-varying dark energy parameters alongside the standard cosmological parameters. Ultimately, we aim to determine whether the high-quality cosmic shear data from HSC-Y3 favors a dynamical dark energy model or remains consistent with a cosmological constant. As we show, our analysis yields results that are broadly consistent with $w = -1$, further cementing the $\Lambda$CDM baseline while providing updated constraints on the $w_0w_a$ parameter space from deep photometric data. 

Our paper is organized as follows. In Sections \ref{Observational Data Used} and \ref{TWO-POINT STATISTICS}, we briefly summarize the HSC-Y3 dataset and the real-space two-point correlation function measurements and the external datasets used. In Sections \ref{THEORETICAL FRAMEWORK or MODEL} and \ref{sec:Modeling_Methodology}, we detail the theoretical framework for the $w_0w_a$CDM model and our treatment of systematic uncertainties. In Section \ref{Results}, we present our primary cosmological constraints and compare them with external datasets, including DESI and DES, and discuss the future scopes in Section \ref{Discussion}. Finally, we conclude in Section \ref{Conclusions}.

\section{Observational Data and Methodology}\label{Observational Data Used}
This section is subdivided into two parts. First, we describe the cosmic shear data that we have used, and then we describe the external datasets that we have used to tighten the constraints on cosmological parameters. 

\subsection{Hyper Suprime Cam - Year 3 Data}\label{Hyper Suprime Cam - Year 3 Data}
We utilize the HSC-Y3 data (S19A release, acquired between March 2014 - April
2019) for this cosmic shear study. In Section \ref{Weak-lensing Shear Catalog},
we discuss the galaxy shear catalog \citet{Li_2022} used for 2PCF measurements.
Section \ref{Photometric Redshift Methodology} contains a description of the
photometric redshift catalog
\citet{2020arXiv200301511N} used for tomographic
binning and distribution inference. Section \ref{Star Catalog for PSF
Systematics} describes the star catalog used to quantify PSF systematics, and
Section \ref{Mock Catalogs} contains description of the mock catalogs used to
determine statistical uncertainties. 

As our analysis builds directly upon the HSC Y3 real-space cosmic shear framework, the following data and methodology descriptions mirror those of \citet{2023PhRvD.108l3518L}. This section is provided to ensure the paper is self-contained. Readers already well-versed in the baseline HSC Y3 analysis may proceed directly to our extended cosmological results in Results Section (\ref{Results}).

\subsubsection{\textbf{Weak-lensing Shear Catalog}}\label{Weak-lensing Shear Catalog}

We utilize the HSC Y3 shape catalog \citep{Li_2022}, which contains galaxy shapes measured from $i$-band wide-field coadded images using the re-Gaussianization algorithm \citep{Hirata_2003}. Following the fiducial analysis, we apply stringent quality cuts to ensure robust shear estimation. This includes removing potential unresolved binary stars, masking a specific region within the GAMA09H field known to suffer from anomalous Point Spread Function (PSF) errors that induce systematic $B$-modes \citep{Zhang_2023}, and filtering out galaxies with highly degenerate photometric redshift posteriors. 

The final, clean catalog covers 416 deg$^2$ across six independent subfields and contains approximately 25 million source galaxies. We fully incorporate the shear calibration framework detailed by \citet{Li_2022}, which corrects for multiplicative, additive, and selection biases using dedicated image simulations. The surveyed area and the effective galaxy number densities ($n_{\text{eff}}$) for each tomographic bin and subfield are summarized in Table \ref{tab:hsc_density}.

\begin{table*}[tbp]
\centering
\caption{The area and effective number density $n_{\text{eff}}^{(i)}$ ($i=1,\dots,4$) \citet{Chang_2013} in each tomographic bin and in six different subfields (i.e., XMM, VVDS, GAMA09H, WIDE12H, GAMA15H, HECTOMAP); and those for the whole HSC-Y3 footprint across the four redshift bins.}
\label{tab:hsc_density}
\begin{tabular}{lcccccc}
\toprule
Fields & Area & $n_{\text{eff}}^{(1)}$ & $n_{\text{eff}}^{(2)}$ & $n_{\text{eff}}^{(3)}$ & $n_{\text{eff}}^{(4)}$ & All \\
 & (deg$^2$) & (arcmin$^{-2}$) & (arcmin$^{-2}$) & (arcmin$^{-2}$) & (arcmin$^{-2}$) & (arcmin$^{-2}$) \\
\midrule
XMM     & 33.17  & 3.44 & 4.46 & 3.66 & 1.94 & 13.51 \\
VVDS    & 96.18  & 3.82 & 5.13 & 4.21 & 2.13 & 15.30 \\
G09H    & 82.36  & 3.99 & 4.74 & 3.81 & 2.11 & 14.65 \\
W12H    & 121.32 & 3.61 & 5.20 & 3.96 & 2.06 & 14.82 \\
G15H    & 40.87  & 3.92 & 5.38 & 4.27 & 2.24 & 15.81 \\
HECT    & 43.06  & 3.74 & 5.34 & 4.05 & 2.32 & 15.44 \\
\midrule
\textbf{All} & \textbf{416.97} & \textbf{3.77} & \textbf{5.07} & \textbf{4.00} & \textbf{2.12} & \textbf{14.96} \\
\bottomrule
\end{tabular}
\end{table*}

\subsubsection{\textbf{Photometric Redshift Methodology}}\label{Photometric Redshift Methodology}

The source galaxies are divided into four tomographic redshift bins spanning $0.3 < z \le 1.5$. The redshift distributions are estimated using a combination of three distinct machine-learning and template-fitting algorithms: dNNz, DEmPz \citep{Hsieh_2014}, and mizuki \citep{Tanaka_2015}, as extensively detailed by \citet{2020arXiv200301511N}. To prevent calibration biases, we explicitly exclude galaxies in the lowest two tomographic bins that exhibit bimodal probability density functions. Such secondary high-redshift peaks extend beyond the calibration range of the CAMIRA-LRG spectroscopic sample \citep{Oguri_2017}, and removing them is critical for reliable cross-correlation calibration \citep{Rau_2023}.

\subsubsection{\textbf{Star Catalog for PSF Systematics}}\label{Star Catalog for PSF Systematics}

To track and marginalize over additive biases introduced by PSF residuals, we rely on the HSC Y3 star catalog \citep{Li_2022}. Specifically, we isolate a high-purity sample of point sources utilized during the coadd modeling phase. As demonstrated by \citet{Zhang_2023}, cross-correlating these "PSF stars" with the galaxy shear catalog provides a statistically powerful framework for explicitly modeling both second- and fourth-order PSF leakages into the observed large-scale two-point correlation functions.

\subsubsection{\textbf{Mock Catalogs}}\label{Mock Catalogs}

The statistical covariance matrix for our two-point correlation functions is derived from a suite of 1404 independent mock shear catalogs. These mocks are generated using full-sky ray-tracing $N$-body simulations \citep{Takahashi_2017} based on a WMAP9 cosmology \citep{Hinshaw_2013}. Following the framework of \citet{Shirasaki_2019}, the simulations are rigorously matched to the exact survey footprint, intrinsic shape noise, and photometric redshift uncertainties of the actual HSC Y3 dataset. This simulation-based approach ensures that non-Gaussian cosmic variance and super-survey covariance effects are naturally incorporated into our error budget.

\subsection{External Datasets}\label{sec:data}

To break the geometric degeneracies inherent in cosmic shear measurements - particularly between the matter density parameter $\Omega_{\rm m}$ and the amplitude of fluctuations $\sigma_8$ - and to provide a lever arm for constraining the time-varying dark energy equation of state parameters ($w_0, w_a$), we combine our HSC Year 3 cosmic shear measurements with external distance probes. We utilise luminosity distance measurements from Type Ia Supernovae (SNIa), angular diameter distance measurements from Baryon Acoustic Oscillations (BAO), supplemented by primordial abundance constraints (constraints on the physical baryon density) from Big Bang Nucleosynthesis (BBN).

\subsubsection{\textbf{Type Ia Supernovae}}\label{Type Ia Supernovae}

Type Ia supernovae serve as standardisable candles that map the expansion history of the Universe via the luminosity distance-redshift relation, $D_L(z)$. We utilise data from three major compilations, each offering distinct advantages in terms of redshift coverage, calibration precision, and sample size. In this work, we draw upon two distinct Dark Energy Survey (DES) supernova compilations for different stages of our analysis. It is crucial to note that these two datasets are \textit{never} combined in the same joint likelihood, strictly avoiding any double-counting of statistical weight.

    For all of our primary joint cosmological constraints (i.e., HSC Y3 + DESI DR2 + SN + BBN), we adopt the DES-Dovekie compilation \citep{2026MNRAS.548ag632P} as our fiducial supernova dataset. We incorporate the DES-Dovekie compilation \citep[e.g.,][]{2026MNRAS.548ag632P}. Building upon the foundational Dark Energy Survey photometric analyses, the Dovekie sample represents a refined dataset of Type Ia supernovae classified utilizing advanced machine learning architectures (such as Supernova) applied directly to survey light curves. Similar to other modern photometric compilations, it bypasses the traditional bottleneck of real-time spectroscopic confirmation, instead relying on high-fidelity host galaxy redshifts and rigorous photometric calibration. Covering a broad and critical redshift range, the Dovekie analysis framework places particular emphasis on forward-modeling the survey selection function and meticulously marginalizing over non-Ia core-collapse contamination using comprehensive mock simulations. By integrating the DES-Dovekie sample, we inject substantial statistical weight into our analysis, exploiting the depth of advanced photometric surveys to deliver robust cosmological constraints that are largely independent of historical spectroscopic targeting biases.

    For comparison purposes, we also consider the photometrically classified SNIa sample from the Dark Energy Survey (DES) 5-year analysis \citep[DES-Y5;][]{2026arXiv260114559D}. We utilize the standard DES 5-year photometrically classified SNIa sample \citep{2026arXiv260114559D} \textit{exclusively} for the external survey consistency checks presented in Section \ref{subsec:des_comparison}. Because the publicly available DES Y5 $3\times2$pt Markov Chain Monte Carlo (MCMC) chains were evaluated using this specific Y5 SN compilation rather than Dovekie, we temporarily substitute Dovekie with the DES-Y5 SN dataset in our own pipeline. This allows us to perform a direct, apples-to-apples consistency comparison between the HSC-anchored and DES-anchored multi-probe baselines. Unlike the JLA \citep{Betoule_2014} and Pantheon+ \citep{Brout_2022, Scolnic_2018} compilations, which rely on spectroscopic confirmation, the DES-Y5 analysis leverages a massive sample of approximately 1600 supernovae identified using the \textsc{DiffImg} difference-imaging pipeline and classified via machine learning algorithms. While lacking individual spectroscopic redshifts for every candidate, the sample benefits from the exquisite photometric calibration of DES and covers a substantial volume in the redshift range $0.025 < z < 0.85$. The analysis rigorously models contamination from core-collapse supernovae and selection effects using large-scale simulations. By including DES-Y5, which covers a wide area, we leverage the statistical power of wide-field photometric surveys, providing complementary constraints that are less sensitive to the specific spectroscopic selection effects dominating older samples, while reducing statistical uncertainties.

\subsubsection{\textbf{Baryon Acoustic Oscillations}}\label{Baryon Acoustic Oscillations}

Baryon Acoustic Oscillations (BAO) imprint a characteristic scale on the clustering of galaxies, providing a standard ruler to constrain the angular diameter distance, $D_A(z)$, and the Hubble parameter, $H(z)$.

    We use the cutting-edge BAO measurements from the Dark Energy Spectroscopic Instrument (DESI) Data Release 2 (DR2). DESI represents a generational leap in spectroscopic volume, utilising 5000 fibre positioners to map the Universe with unprecedented density. The DR2 analysis \citep{desicollaboration2025desidr2resultsii, 2025PhRvD.112h3515A} derives BAO constraints from over 14 million extragalactic sources, categorised into distinct tracers: the Bright Galaxy Survey (BGS) at $z \sim 0.3$, Luminous Red Galaxies (LRGs) at $z \sim 0.5-0.8$, Emission Line Galaxies (ELGs) extending to $z \sim 1.6$, and Quasars (QSO) probing up to $z \sim 2.1$. Additionally, it includes constraints from the Lyman-$\alpha$ forest seen in high-redshift quasar spectra, pushing the lever arm to $z > 2.1$. This multi-tracer approach allows us to probe the evolution of the expansion history $H(z)$ continuously from the local Universe out to the matter-dominated epoch, providing critical sensitivity to dynamical dark energy models (e.g., $w_0 w_a$CDM) that may deviate from $\Lambda$CDM at high redshifts. These data provide constraints on $D_V(z)/r_d$ over a wide redshift range up to $z \sim 3.5$, offering critical leverage for constraining the high-redshift behaviour of dark energy.

\subsubsection{\textbf{Big Bang Nucleosynthesis}}\label{Big Bang Nucleosynthesis}

Big Bang Nucleosynthesis (BBN) provides a pristine probe of the early Universe, anchoring the physical baryon density, $\omega_b \equiv \Omega_b h^2$, independently of late-time cosmological probes or the Cosmic Microwave Background. To tightly constrain the baryon density, we incorporate BBN likelihoods based on high-resolution spectroscopic measurements of primordial deuterium abundances in unevolved damped Lyman-$\alpha$ systems. Specifically, we adopt the benchmark measurement of the primordial deuterium-to-hydrogen ratio, $(\text{D/H})_P = (2.527 \pm 0.030) \times 10^{-5}$, from \citet{Cooke_2018}. By comparing these astrophysical observations with theoretical BBN predictions—while carefully accounting for uncertainties in standard nuclear reaction rates—this dataset provides a stringent, a complementary anchor for $\omega_b$ that efficiently breaks parameter degeneracies in our joint cosmological analysis.

\section{TWO-POINT STATISTICS}\label{TWO-POINT STATISTICS}

We utilize the galaxy shear two-point correlation functions (2PCFs), denoted as $\xi_\pm(\theta)$, as the primary observables for obtaining our cosmological constraints \citet{BARTELMANN2001291}. Because the measurement of the data vector, the construction of the covariance matrix, and the systematic $B$-mode validation tests are identical to those of the fiducial HSC-Y3 real-space analysis, we provide only a brief overview here. We refer the reader to \citet{2023PhRvD.108l3518L} for the exhaustive measurement methodology.

\subsection{Two-Point Correlation Functions}\label{Two-Point Correlation Functions}

\begin{figure}[htbp]
\centering
\includegraphics[width=\columnwidth]{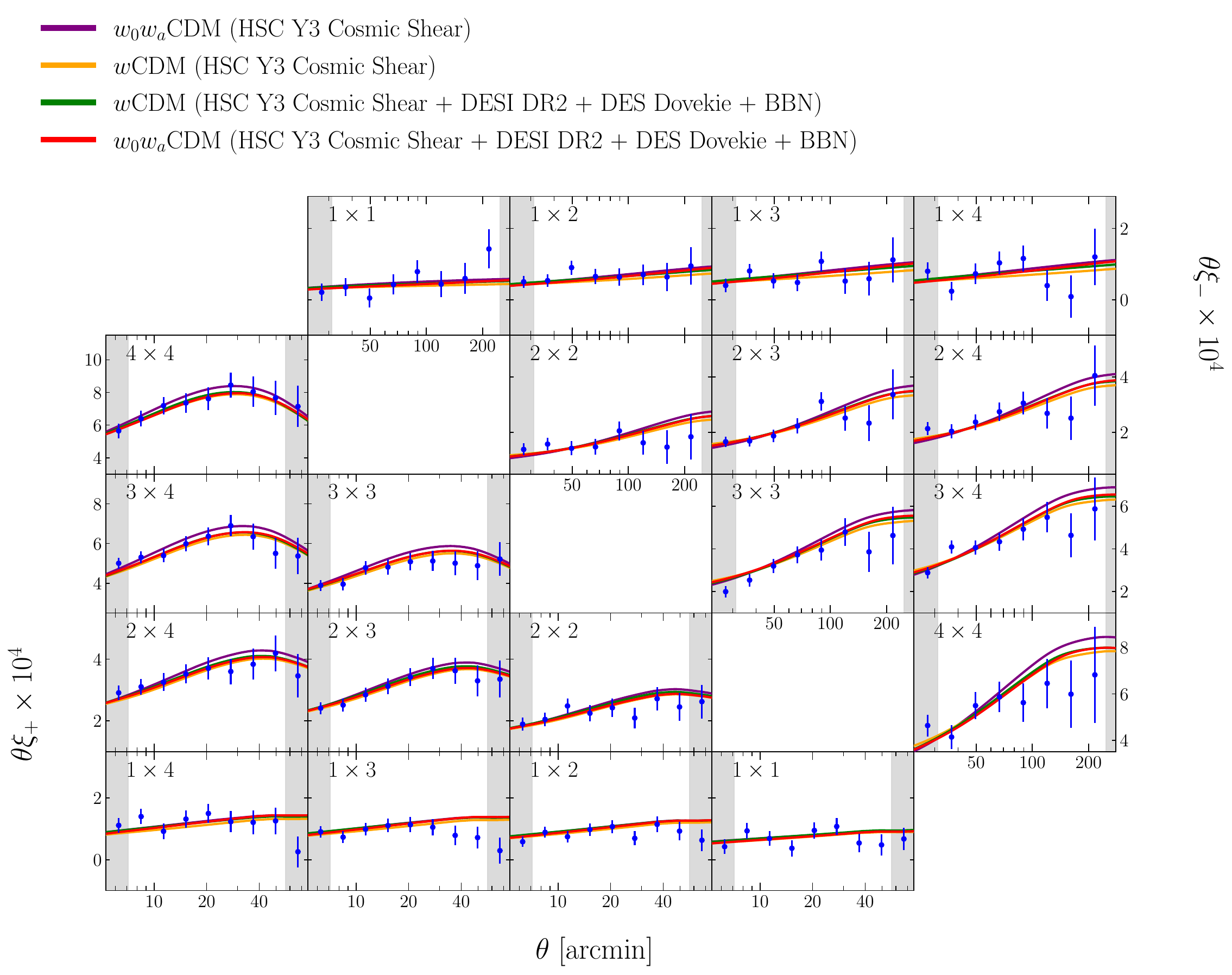}
\caption{ The ten 2PCFs including four autocorrelations and six cross-correlations between the four tomographic redshift bins
(labeled with 1–4). This plot shows the 2PCFs on scales $5.3 < \xi_+ < 76$ [arcmin] for $\xi_+$, and $23.2 < \xi_- < 248$ [arcmin] for $\xi_-$. The unshaded region refers to the fiducial scale cut: $7.1 < \xi_+ < 56.6$ [arcmin] for $\xi_+$, and $31.2 < \xi_- < 248$ [arcmin] for $\xi_-$ . The errorbars are estimated with mock catalogs. The solid lines show the model predictions at the maximum-posterior (MAP) sample of each chain, as discussed in later Section \ref{subsec:best_fit_2pcf}.}
\label{fig: bestfit}
\end{figure}

\noindent We quantify the cosmic shear signal by computing the standard auto- and cross-correlation functions, $\xi_{\pm}(\theta)$, across our four tomographic bins using the \texttt{TreeCorr}\footnote[1]{ https://github.com/rmjarvis/TreeCorr} software \citep{Jarvis_2004, Porth_2024}. The measurements are derived directly from the calibrated galaxy shape catalog and are binned in equal logarithmic angular intervals.

To ensure robust cosmological constraints, we apply strict angular scale cuts to the data vector. Following \citet{2023PhRvD.108l3518L}, we adopt fixed angular cuts across all redshift bins to isolate the cosmological signal from known systematics. Specifically, we exclude small scales ($\theta < 7.1$ arcmin for $\xi_+$, $\theta < 31.2$ arcmin for $\xi_-$) to minimize modeling uncertainties driven by nonlinear baryonic physics, and we exclude large scales ($\theta > 56.6$ arcmin for $\xi_+$, $\theta > 248$ arcmin for $\xi_-$) to avoid contamination from systematic $B$-modes. 

Fig. \ref{fig: bestfit} shows the measured 2PCFs from the galaxy shear catalog, with the unshaded regions denoting the scales used for our fiducial analysis. The resulting data vector comprises 7 angular bins for both $\xi_+$ and $\xi_-$ across all 10 tomographic combinations (4 auto-correlations and 6 cross-correlations), yielding a total of 140 data points. We note that a parallel Fourier-space analysis of this same catalog is presented in \citet{Dalal_2023}.

\subsection{Covariance Matrix Estimation}\label{Covariance Matrix Estimation}

In this work, we adopt the exact same covariance matrix, $\mathbf{C}$, utilized in the baseline HSC Y3 $\Lambda$CDM cosmic shear analysis \citep{2023PhRvD.108l3518L}. 

Briefly, the covariance matrix is evaluated directly from the sample variance of 1404 mock shear catalogs based on full-sky ray-traced $N$-body simulations \citep{Takahashi_2017}. This simulation-based approach naturally accounts for non-Gaussian cosmic variance arising from non-linear structure formation, super-survey covariance \citep{Takada_2013, Shirasaki_2019}, and observational shape noise properly matched to the HSC survey geometry \citep{Li_2022}. We retain the standard assumption of evaluating the covariance at a fixed fiducial cosmology, as neglecting the cosmology dependence of the covariance matrix has been shown to introduce negligible bias ($\le 1\%$ of the statistical error) to parameter constraints in current weak lensing surveys \citep{Kodwani_2019}.

For an exhaustive description of the mock catalog generation, simulation resolution corrections, and the statistical validation of the covariance matrix, we direct the reader to \citet{2023PhRvD.108l3518L}.

\subsection{\texorpdfstring{$B$}{B}-mode Decomposition and Systematics}\label{$B$-mode Decomposition and Systematics}

The measured correlation functions, $\xi_\pm$, contain contributions from both the gradient (curl-free $E$-mode) and curl ($B$-mode) components of the shear field. Because physical astrophysical sources of $B$-modes \citep[e.g.,][]{Schneider_2002, Krause_2010, Blazek_2019} are expected to be negligible compared to the $E$-mode signal, the detection of a significant $B$-mode serves as a standard diagnostic for systematic errors in the data.

In this work, we do not independently re-evaluate the $B$-mode decomposition. Instead, we rely on the extensive null tests performed on this catalog by \citet{2023PhRvD.108l3518L}. By separating the $E$- and $B$-modes following the integral relations derived by \citet{Schneider_2002}, \citet{2023PhRvD.108l3518L} identified significant $B$-mode contamination in $\xi_+$ at large angular scales, particularly in the high-redshift bins. 

To ensure our cosmological inference remains unbiased by these systematics, we adopt their validated scale cuts. Specifically, we discard $\xi_+$ measurements at $\theta \ge 56.6$ arcmin. For $\xi_-$, we apply a conservative cut at $\theta \ge 248$ arcmin; while statistically significant $B$-modes were not detected here, these scales are dominated by cosmic variance and contribute negligibly to the overall signal-to-noise ratio. (Note that small-scale cuts are determined separately by baryonic modeling requirements, as discussed in Section \ref{subsec:inference_validation}. 

As demonstrated by \citet{2023PhRvD.108l3518L}, post-masking, the probability that the residual $B$-modes in the fiducial data vector are consistent with zero is $p=0.143$ for $\xi_+$ and $p=0.237$ for $\xi_-$, confirming the robustness of these scale cuts for our analysis.

\begin{table}[htbp]
\centering
\caption{Model parameters and priors used in our fiducial cosmological parameter inference. The label $\mathcal{U}(a, b)$ denotes a noninformative flat prior between $a$ and $b$, and $\mathcal{N}(\mu, \sigma)$ denotes a normal distribution with mean $\mu$ and width $\sigma$.}
\label{tab:priors}
\begin{tabular}{lc}
\toprule
Parameter & Prior \\
\midrule
\multicolumn{2}{l}{\textit{Cosmological parameters (\ref{subsec:cosmo_astro_modeling})}} \\
\midrule
$\Omega_{\rm m}$ & $\mathcal{U}(0.1, 0.7)$ \\
$A_s \, (\times 10^{-9})$ & $\mathcal{U}(0.5, 10)$ \\
$n_s$ & $\mathcal{U}(0.87, 1.07)$ \\
$h_0$ & $\mathcal{U}(0.62, 0.80)$ \\
$\omega_b$ & $\mathcal{U}(0.02, 0.025)$ \\
\midrule
\multicolumn{2}{l}{\textit{Dark energy parameters (\ref{subsec:working_pipeline})}} \\
\midrule
$w$ ($w$CDM) & $\mathcal{U}(-2.0, -0.5)$ \\
$w_0$ ($w_0w_a$CDM) & $\mathcal{U}(-2.0, -0.5)$ \\
$w_a$ ($w_0w_a$CDM) & $\mathcal{U}(-2.5, 2.0)$ \\
\midrule
\multicolumn{2}{l}{\textit{Baryonic feedback parameters (\ref{subsec:cosmo_astro_modeling})}} \\
\midrule
$A_b$ & $\mathcal{U}(2, 3.13)$ \\
\midrule
\multicolumn{2}{l}{\textit{Intrinsic alignment parameters (\ref{subsec:cosmo_astro_modeling})}} \\
\midrule
$A_1$ & $\mathcal{U}(-6, 6)$ \\
$\eta_1$ & $\mathcal{U}(-6, 6)$ \\
$A_2$ & $\mathcal{U}(-6, 6)$ \\
$\eta_2$ & $\mathcal{U}(-6, 6)$ \\
$b_{ta}$ & $\mathcal{U}(0, 2)$ \\
\midrule
\multicolumn{2}{l}{\textit{Photo-$z$ systematics (\ref{subsec:observational_systematics})}} \\
\midrule
$\Delta z_1$ & $\mathcal{N}(0, 0.024)$ \\
$\Delta z_2$ & $\mathcal{N}(0, 0.022)$ \\
$\Delta z_3$ & $\mathcal{U}(-1, 1)$ \\
$\Delta z_4$ & $\mathcal{U}(-1, 1)$ \\
\midrule
\multicolumn{2}{l}{\textit{Shear calibration biases (\ref{subsec:observational_systematics})}} \\
\midrule
$\Delta m_1$ & $\mathcal{N}(0.0, 0.01)$ \\
$\Delta m_2$ & $\mathcal{N}(0.0, 0.01)$ \\
$\Delta m_3$ & $\mathcal{N}(0.0, 0.01)$ \\
$\Delta m_4$ & $\mathcal{N}(0.0, 0.01)$ \\
\midrule
\multicolumn{2}{l}{\textit{PSF systematics (\ref{subsec:observational_systematics})}} \\
\midrule
$\alpha'^{(2)}$ & $\mathcal{N}(0, 1)$ \\
$\beta'^{(2)}$ & $\mathcal{N}(0, 1)$ \\
$\alpha'^{(4)}$ & $\mathcal{N}(0, 1)$ \\
$\beta'^{(4)}$ & $\mathcal{N}(0, 1)$ \\
\bottomrule
\end{tabular}
\end{table}

\section{THEORETICAL FRAMEWORK or MODEL}\label{THEORETICAL FRAMEWORK or MODEL}

In this work, we adopt the forward-modeling pipeline developed for the baseline HSC Y3 $\Lambda$CDM cosmic shear analysis by \citet{2023PhRvD.108l3518L} and extend it to evaluate dynamical dark energy models ($w$CDM and $w_0w_a$CDM). Our model is implemented within the CosmoSIS framework \citep{Zuntz_2015} and projects the 3D matter power spectrum into the observable real-space tomographic two-point correlation functions (2PCFs), $\xi^{ij}_\pm(\theta)$, using the Limber approximation and the flat-sky Hankel transform. 

Because the foundational methodology, parameterization, and system testing were comprehensively established in the baseline analysis, we provide only a brief summary of the physical and systematic components here. We refer the reader to \citet{2023PhRvD.108l3518L} and the parallel Fourier-space analysis by \citet{Dalal_2023} for the complete mathematical formalism and prior definitions.

\subsection{Cosmological and Astrophysical Modeling}
\label{subsec:cosmo_astro_modeling}

To evaluate the nonlinear matter power spectrum across our extended cosmological parameter space, we utilize the BACCO emulator (v1.0.0) \citep{Arico__2022}, which provides accurate predictions for the linear transfer function and natively supports time-varying dark energy equations of state. For the $w$CDM model, we compute the linear matter power spectrum with the BACCO emulator (v1.0.0) \citep{Arico__2022}. For the $w_0w_a$CDM model, whose prior volume extends beyond the training range of the emulator, we instead compute the linear matter power spectrum directly with the Einstein-Boltzmann solver CAMB \citep{Lewis_2000}. We account for small-scale baryonic feedback using the HMCode2016 halo-model framework \citep{Mead_2016}, marginalizing over the feedback amplitude $A_b$.

The intrinsic alignment (IA) of galaxies—which contaminates the pure gravitational lensing signal—is modeled using the Tidal Alignment and Tidal Torque (TATT) framework \citep{Blazek_2019}. TATT expands the alignment field to quadratic order in the density and tidal fields. Following the model selection tests in \citet{2023PhRvD.108l3518L}, we adopt TATT over the simpler Nonlinear Alignment (NLA) model \citep{Bridle_2007, Joachimi_2011, Hirata_2007, 2004PhRvD..70f3526H} as our fiducial choice, as it provides a more complete physical description of IA without degrading cosmological constraining power.

\subsection{Observational Systematics}
\label{subsec:observational_systematics}

Our inference pipeline explicitly marginalizes over three primary sources of observational systematics, introducing 12 nuisance parameters. The first set of four parameters concern photometric redshift uncertainties. We mode the errors in the source redshift distribution as mean shifts, $\Delta z_i$. Following \citet{Zhang_2022}, we use informative Gaussian priors for the low-redshift bins (Bins 1 and 2). However, to avoid biasing our cosmology due to a lack of spectroscopic calibration at high redshifts, we adopt conservative flat priors for Bins 3 and 4. The next set of four parameters are related to residual multiplicative biases in shear calibration in each redshift bin, parameterized by $\Delta m_i$. These are constrained by Gaussian priors of $\mathcal{N}(0, 0.01)$ based on extensive image simulations \citep{Li_2022}. Finally, we also have four parameters to account for systematics due to PSF leakage. Following \citet{Zhang_2023}, we account for leakage and modeling errors from both the second-order and the spin-2 component of the fourth-order PSF moments, applying the necessary additive corrections directly to the predicted $\xi_+$ signal.

\subsection{Bayesian Inference and Model Validation}
\label{subsec:inference_validation}

We perform Bayesian parameter estimation to sample the high-dimensional posterior distribution (combining cosmological, baryonic, IA, and systematic parameters) using the PolyChord nested sampler \citep{Handley_2015}. For computational efficiency during internal consistency tests, we also utilize MultiNest \citep{Feroz_2009}, relying on the covariance matrix derived from the 1404 mock catalogs described in Section \ref{Covariance Matrix Estimation}.

The robustness of this exact modeling pipeline was rigorously validated by \citet{2023PhRvD.108l3518L} using synthetic data vectors contaminated with alternative physical prescriptions (e.g., OWLS-AGN baryonic feedback and the MiraTitan II emulator). Their Maximum A Posteriori (MAP) optimizations demonstrated that systematic biases in the $S_8$ parameter remain well below the statistical error ($< 0.15\sigma$), satisfying the standard $< 0.3\sigma$ threshold for 2D posterior bias. While projection volume effects during marginalization can induce artificial shifts in the matter density $\Omega_{\rm m}$ \citep{Joachimi_2021}, $S_8$ remains robust. We rely on these validation results for our extended cosmological analysis, ensuring that our constraints on $S_8$ and the dark energy equation of state ($w_0, w_a$) are not driven by unmodeled baryonic or systematic artifacts.

\section{Modeling Methodology}\label{sec:Modeling_Methodology} 

\subsection{CosmoSIS: The COSMOlogical Survey Inference System}
For our statistical analysis, we utilize the COSMOlogical Survey Inference System \citep[CosmoSIS;][]{Zuntz_2015, noauthor_undated-in}. CosmoSIS is a flexible, modular framework designed to streamline cosmological parameter estimation. By structuring analytical steps into independent, interchangeable calculation modules, the software prioritizes code reusability, verifiability, and robust debugging. Crucially, it acts as a centralized pipeline that seamlessly integrates established community codes to predict theoretical observables. This architecture provides a unified, accessible environment for evaluating experimental likelihoods and exploring complex parameter spaces using a variety of advanced statistical sampling techniques.

\subsection{Parameter Inference Pipeline}
\label{subsec:working_pipeline}

We adopt the fiducial parameter inference pipeline from the HSC Year 3 real-space cosmic shear analysis \citep{2023PhRvD.108l3518L}, implemented within the \texttt{CosmoSIS} framework. To extend the baseline $\Lambda$CDM model, we introduce the dark energy equation of state $w$ as a free parameter. We first evaluate the $w$CDM model by varying a constant equation of state, $w$. Subsequently, we evaluate the dynamical $w_0w_a$CDM model utilizing the standard Chevallier-Polarski-Linder (CPL) parameterization, $w(a) = w_0 + (1-a)w_a$, sampling both $w_0$ and $w_a$ subject to the priors summarized in Table \ref{tab:priors}.

\section{Results}\label{Results}

In this section, we present the cosmological constraints derived from the Hyper Suprime-Cam (HSC) Year 3 real-space cosmic shear data, extending the previous baseline $\Lambda$CDM analysis \citep{2023PhRvD.108l3518L} to dynamical dark energy models. 
To break the geometric degeneracies inherent to weak lensing, we combine our HSC Y3 data with DESI DR2 BAO, the recalibrated DES five-year SN sample (DOVEKIE), and Big Bang Nucleosynthesis (BBN) priors, establishing a constraining, fully CMB-independent joint dataset.

\subsection{Maximum-Posterior Two-Point Correlation Functions}
\label{subsec:best_fit_2pcf}

We first verify that our extended cosmological models provide a good description of the observed data vectors. Figure~\ref{fig: bestfit} shows the measured $\xi_+$ and $\xi_-$ two-point correlation functions for all ten tomographic bin combinations, together with the model predictions for the $w$CDM and $w_0w_a$CDM models. For each case, the model is evaluated at the maximum-posterior (MAP) sample of the corresponding chain, using the parameter values of that sample as input to our CosmoSIS pipeline. We show the MAP models for both the HSC-only analyses and the joint analyses (HSC Y3 + DESI DR2 + DES DOVEKIE + BBN). In all cases, the models reproduce the measured correlation functions across the angular scales used in our fiducial analysis. The $w$CDM and $w_0w_a$CDM predictions are nearly indistinguishable within the data uncertainties, indicating that the additional dark energy parameters neither degrade nor substantially improve the fit to the HSC shear data (Figure \ref{fig: bestfit}).

\subsection{Constraints on \texorpdfstring{$w$}{w}CDM}
\label{subsec:wcdm_constraints}

\begin{figure}[htbp]
\centering
\includegraphics[width=\columnwidth]{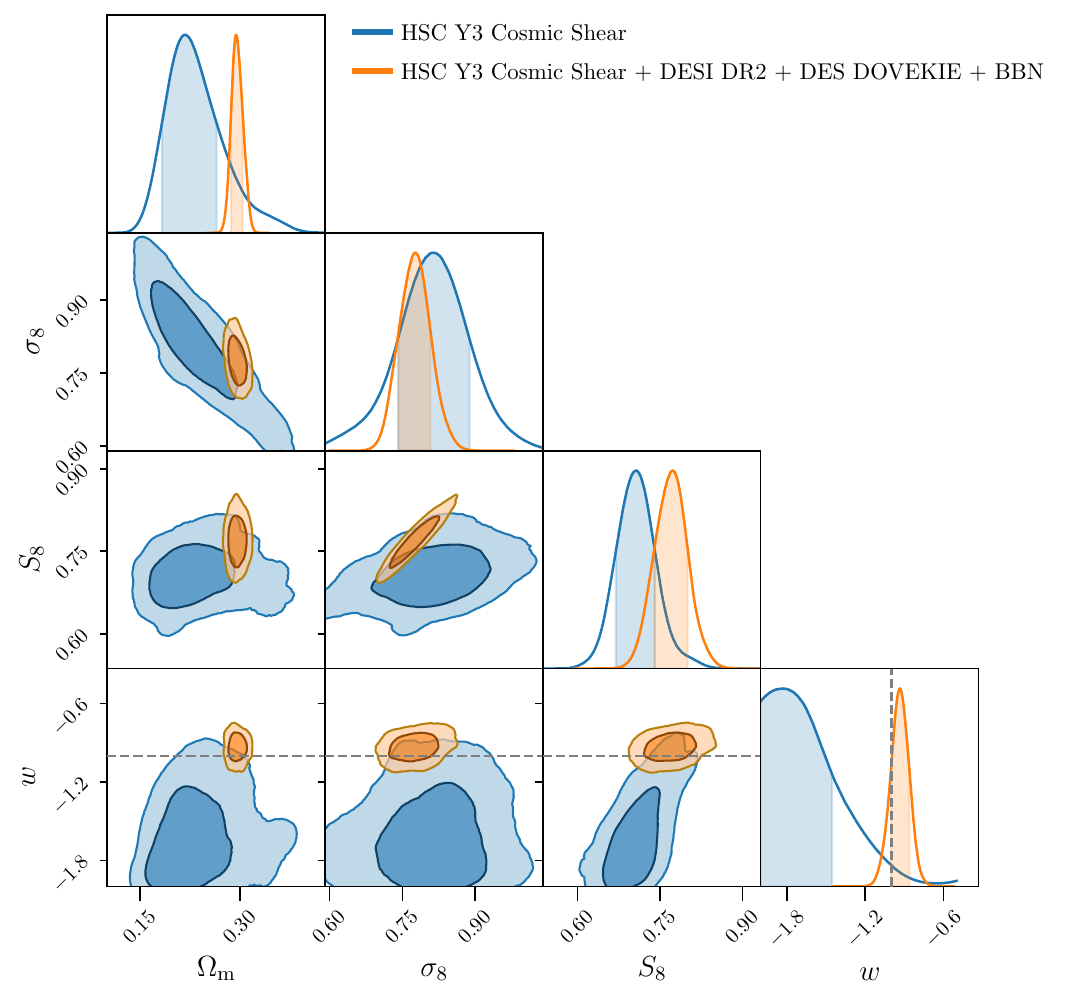}
\caption{Constraints on $\Omega_{\rm m}$, $S_8$ and dark energy equation of state parameter $w$ using Hyper-Suprime Cam Year 3 data. The different colors are for different likelihoods used to constrain the parameters and are mentioned in the legend.}
\label{fig: wCDM result 1}
\end{figure}

Allowing the dark energy equation of state $w$ to vary as a free constant introduces a well-known geometric degeneracy when relying solely on cosmic shear data. As seen in Figure \ref{fig: wCDM result 1}, the HSC-only analysis (blue contours) yields an elongated posterior in the $\Omega_{\rm m} - w$ plane. Cosmic shear alone barely constrains dark energy, and we find that this resulting $w - S_8$ degeneracy pulls the marginalized $S_8$ value somewhat lower to $S_8 = 0.709^{+0.037}_{-0.036}$, than its $\Lambda$CDM counterpart, for which the HSC-Y3 cosmic shear analyses found $S_8 = 0.769^{+0.031}_{-0.034}$ from the real-space two-point correlation functions \citep{2023PhRvD.108l3518L} and $S_8 = 0.776^{+0.032}_{-0.033}$ from the Fourier-space power spectra \citep{Dalal_2023}. This is primarily driven by the lower value of $\Omega_{\rm m}$.

This motivates the inclusion of our CMB-independent background probes, which can break this degeneracy (Figure \ref{fig: wCDM result 1}). The joint analysis (orange contours) tightens the constraints, yielding $w = -0.936^{+0.074}_{-0.067}$, which remains statistically consistent with a cosmological constant. Importantly, once the background geometry breaks the parameter degeneracy, the joint analysis yields $S_8 = 0.774^{+0.027}_{-0.034}$. This successfully recovers the HSC-Y3 $\Lambda$CDM value, demonstrating that the weak lensing amplitude is robust to constant dark energy equation of state extension. The extended parameter space, shown in Figure \ref{fig: wCDM result Appendix} in Appendix \ref{2D posteriors for internal tests}, confirms that the intrinsic alignment amplitude ($A_I$) and photometric redshift calibration uncertainties ($\Delta z$) remain stable and are also unaffected by the external data. 

\begin{figure}[htbp]
\centering
\includegraphics[width=\columnwidth]{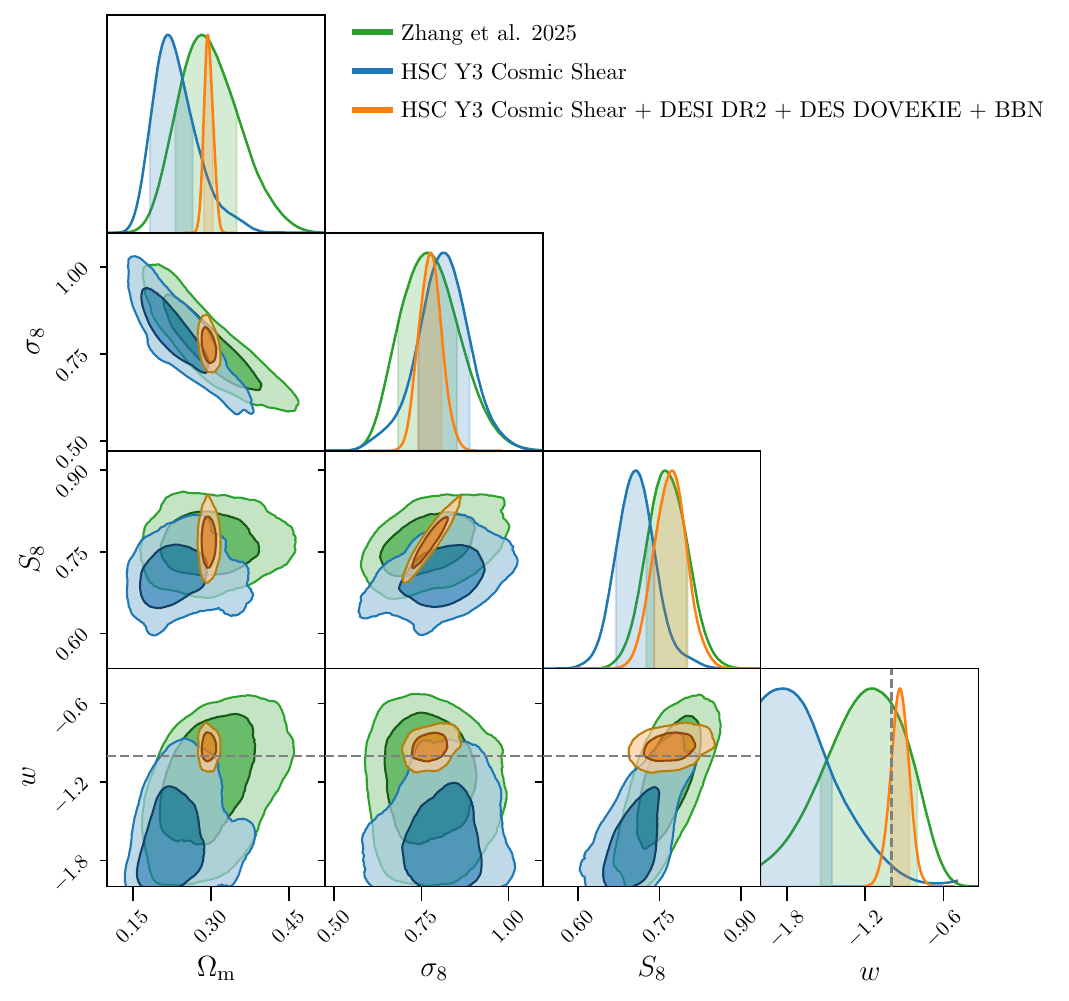}
\caption{Constraints on $\Omega_{\rm m}$, $S_8$ and dark energy equation of state parameter $w$ using Hyper-Suprime Cam Year 3 data. The different colors are for different likelihoods used to constrain the parameters and are mentioned in the legend and compared with the results from \citet{2026PhRvD.113j3530Z}}
\label{fig: wCDM result 2}
\end{figure}

We also compare our $w$CDM constraints with recent literature. As shown in Figure \ref{fig: wCDM result 2}, our joint constraints are in strong agreement with the analysis presented in \citet{2026PhRvD.113j3530Z}. This consistency is largely expected given that both analyses share the HSC Y3 source galaxy shear catalog. However, the data vector combinations and modeling frameworks differ fundamentally. \citet{2026PhRvD.113j3530Z} performed a $3\times2$pt analysis, combining HSC Y3 cosmic shear with galaxy-galaxy lensing and galaxy clustering using an SDSS lens sample, and employed a point-mass correction model to extract information from small-scale lensing signals (down to $2 h^{-1}\mathrm{Mpc}$). In contrast, our methodology isolates the large-scale cosmic shear signal to anchor the amplitude of structure growth, relying strictly on completely independent, background distance measurements (DESI DR2 BAO and DES DOVEKIE supernovae) to constrain the expansion history. These distinct approaches, one which integrates the low-redshift large-scale structure cross-correlations and small-scale modeling, and the other which leverages purely geometric, background probes, both converge in the $\Omega_{\rm m} - S_8$ plane. This validates the robustness of the inferences.

\begin{figure}[htbp]
\centering
\includegraphics[width=\columnwidth]{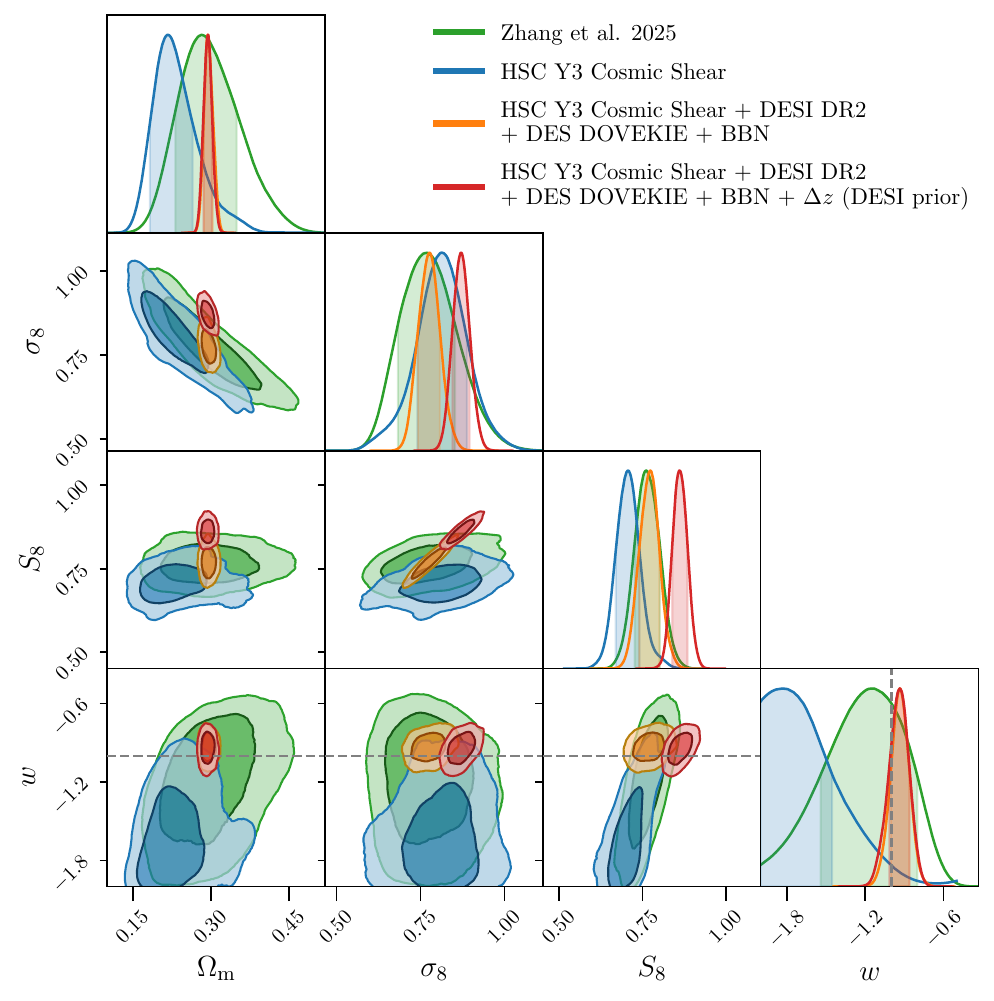}
\caption{Constraints on $\Omega_{\rm m}$, $S_8$ and dark energy equation of state parameter $w$ using Hyper-Suprime Cam Year 3 data. The different colors are for different likelihoods used to constrain the parameters and are mentioned in the legend and compared with the results from \citet{2026PhRvD.113j3530Z}}
\label{fig: wCDM result desi prior}
\end{figure}

Figure \ref{fig: wCDM result desi prior} displays the 1D and 2D marginalized posteriors for the $w$CDM parameter space ($\Omega_{\rm m}, \sigma_8, S_8, w$). As expected, HSC Y3 cosmic shear alone (orange contours) exhibits broad constraints on the dark energy equation of state ($w$) and a strong geometric degeneracy in the $\Omega_{\rm m}$--$\sigma_8$ plane. The addition of external background probes (DES DOVEKIE supernovae, DESI DR2 BAO, and BBN) breaks these degeneracies, reducing the uncertainties by an order of magnitude (green contours).   Next, we investigate the sensitivity of these joint constraints to the photometric redshift calibration of the source galaxies. Our baseline joint analysis uses conservative, uninformative flat priors on the redshift biases ($\Delta z$). When we replace these with the informative, tight cross-correlation priors derived from DESI data \citep{Choppin_de_Janvry_2026} (red contours), we observe a significant, systematic shift in the posterior. First, the DESI redshift priors pull the structure growth parameter $S_8$ to a higher value, bringing our cosmic shear constraints into strong agreement with the independent HSC Y3 + SDSS 3x2pt analysis by \citet{2026PhRvD.113j3530Z} (blue contours). 

\subsection{Constraints on \texorpdfstring{$w_0w_a$}{w0wa}CDM}
\label{subsec:w0wacdm_constraints}

\begin{figure}[htbp]
\centering
\includegraphics[width=\columnwidth]{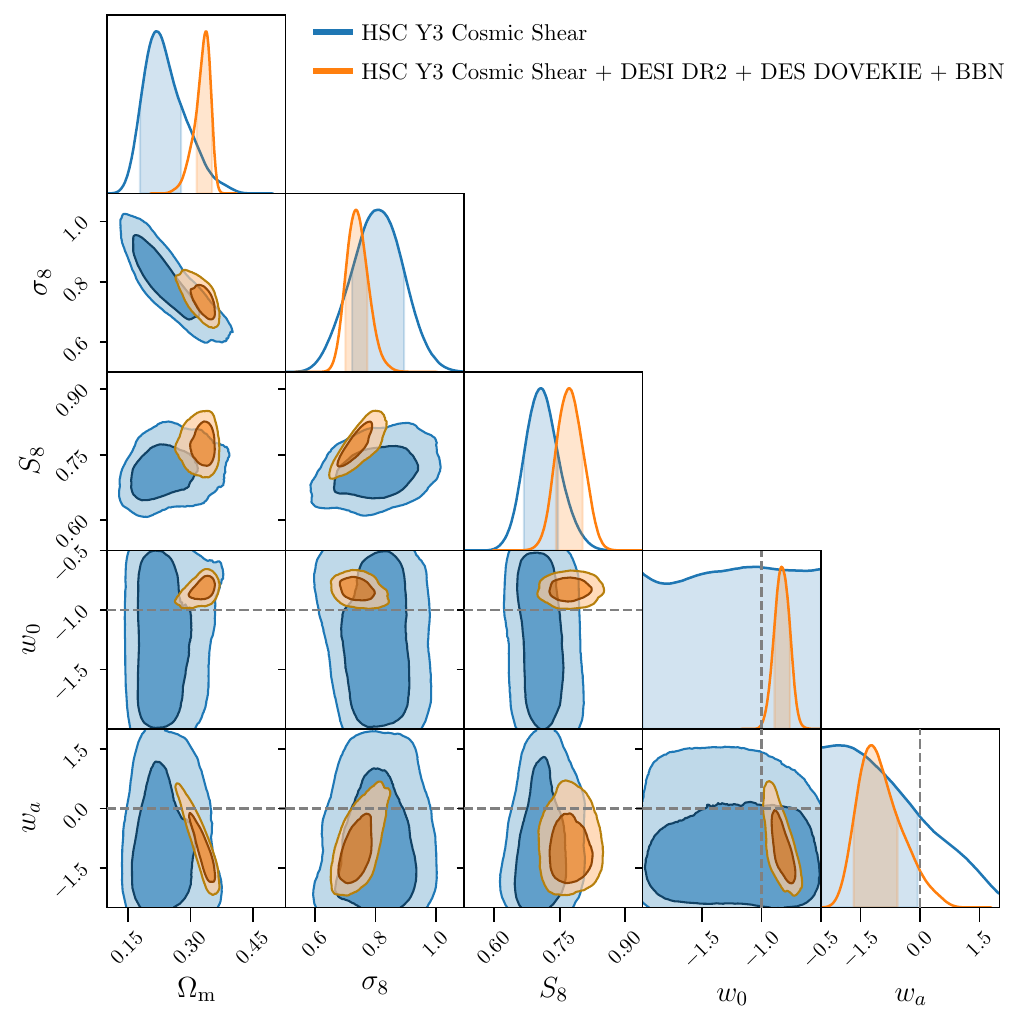}
\caption{Constraints on $\Omega_{\rm m}$, $S_8$ dark energy equation of state parameters $w_0$ and $w_a$, using Hyper-Suprime Cam Year 3 data. The different colors are for different likelihoods used to constrain the parameters and are mentioned in the legend.}
\label{fig: w0waCDM result 1}
\end{figure}

We next generalize to a time-varying dark energy equation of state. The marginalized posteriors for this model are shown in Figure \ref{fig: w0waCDM result 1}. Similar to the $w$CDM case, the HSC Y3 cosmic shear data alone cannot place meaningful bounds on the evolutionary parameter $w_a$, resulting in a completely unconstrained posterior that spans the entirety of the prior volume. 

Once again we combine these late-time growth measurements (HSC) with precise expansion history data (DESI and DES SN). The joint likelihood (orange contours) successfully localizes the dark energy parameters, yielding $w_0 = -0.832^{+0.071}_{-0.059}$ and $w_a = -1.23^{+0.66}_{-0.44}$ (Figure \ref{fig: w0waCDM result 1}). This represents a $2.4\sigma$ preference for dynamical dark energy, consistent with recent DESI findings. Despite this shift in the background cosmology, the joint constraint on structure growth remains $S_8 = 0.772^{+0.031}_{-0.030}$. This confirms that the HSC-Y3 lensing amplitude is fundamentally robust to the assumed dark energy model, and that the observed dynamical dark energy signal is currently primarily driven by the geometric probes. The nuisance parameter posteriors (Figure \ref{fig: w0waCDM result Appendix} in \ref{2D posteriors for internal tests}) exhibit no unexpected oddities, and confirm that our systematics mitigation remains fully effective in extended cosmologies.

\subsection{Consistency with External Lensing Surveys}
\label{subsec:des_comparison}

\begin{figure}[htbp]
\centering
\includegraphics[width=\columnwidth]{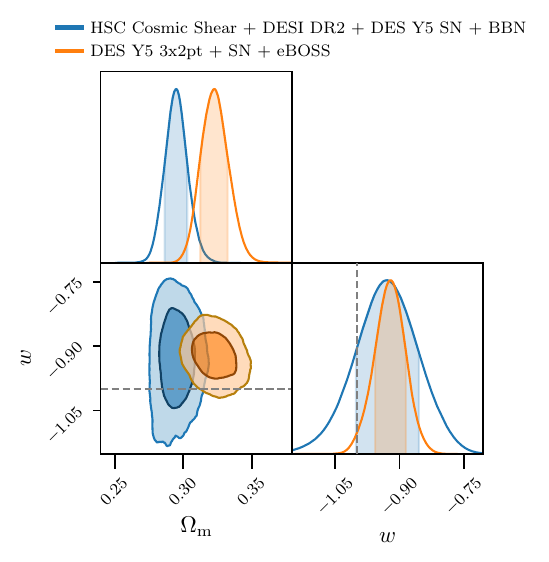}
\caption{Constraints on $\Omega_{\rm m}$ and dark energy equation of state parameter $w$ using Hyper-Suprime Cam Year 3 data. The different colors are for different likelihoods used to constrain the parameters and are mentioned in the legend and compared with the results from DES Y5 3$\times$2 pt.}
\label{fig: wCDM result 3}
\end{figure}

We compare the constraining power of our results in our joint baseline (HSC Cosmic Shear + DESI DR2 + DES SN + BBN) against the recent DES Y5 $3\times2$pt + SN + eBOSS analysis \citep{2026arXiv260114559D, Abbott_2022, 2021PhRvD.103h3533A}. In the $w$CDM paradigm (Figure \ref{fig: wCDM result 3}), both surveys yield consistent marginalized posteriors. The constraints on $\Omega_{\rm m}$ and $w$ show excellent agreement. 

\begin{figure}[htbp]
\centering
\includegraphics[width=\columnwidth]{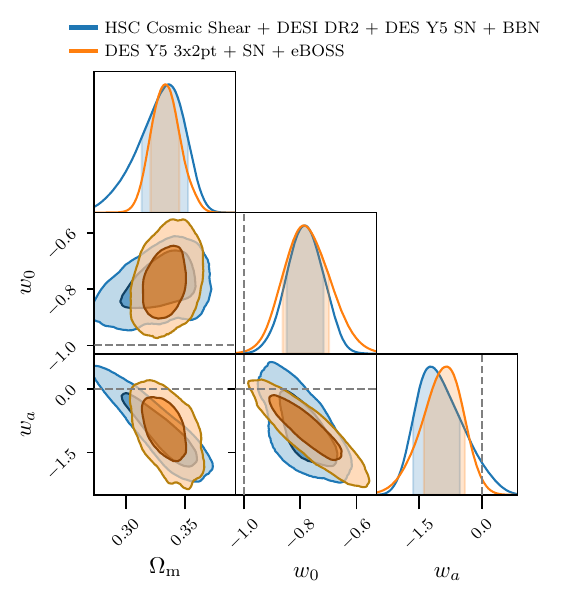}
\caption{Constraints on $\Omega_{\rm m}$ and dark energy equation of state parameters $w_0$ and $w_a$ using Hyper-Suprime Cam Year 3 data. The different colors are for different likelihoods used to constrain the parameters and are mentioned in the legend and compared with the results from DES Y5 3$\times$2 pt.}
\label{fig: w0waCDM result 2}
\end{figure}

This excellent agreement extends into the $w_0w_a$CDM parameter space, as illustrated in Figure \ref{fig: w0waCDM result 2}. Both independent lensing surveys, when combined with their respective external datasets, trace a consistent degeneracy direction that favors $w_0 > -1$ and $w_a < 0$. The agreement across independent observing campaigns, telescope systematics, and varying analysis choices bolsters confidence in the global constraints on dynamical dark energy and serves as a key consistency test ahead of next-generation surveys like Rubin LSST and Euclid.
     
\subsection{Internal Robustness Checks}
\label{subsec:internal_consistency}

\begin{figure}[htbp]
\centering
\includegraphics[width=\columnwidth]{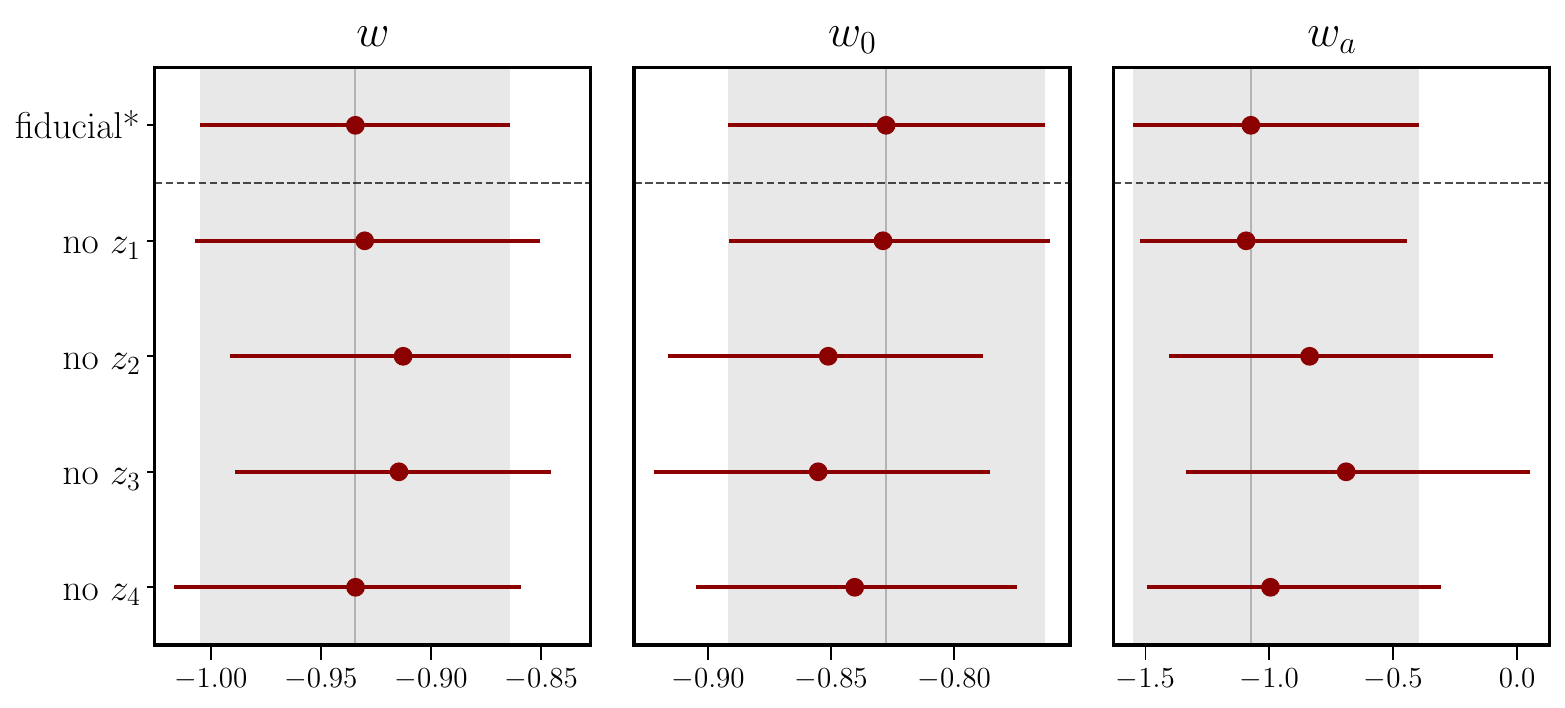}
\caption{The 68\% CIs of 1D projected modes on the dark energy equation of state parameters $w$, and $w_0$, $w_a$ for different splits of the HSC-Y3 data. We have two groups divided by horizontal dashed lines. The first group is the fiducial setup (HSC Y3 Cosmic Shear + DESI DR2 + DES DOVEKIE + BBN) while the second group is for removal of different redshift bins. All results are sampled with PolyChord.}
\label{fig: combined wcdm and w0wacdm whisker plot}
\end{figure}

To ensure our constraints are not driven by localized systematics in specific redshift intervals, we perform a leave-one-out robustness test on our tomographic bins. Figure \ref{fig: combined wcdm and w0wacdm whisker plot} presents the 1D marginalized modes and 68\% confidence intervals for the dark energy equation of state parameters ($w$ for the $w$CDM model; $w_0, w_a$ for the $w_0w_a$CDM model) when sequentially removing tomographic bins $z_1$ through $z_4$ from the joint analysis.     
The time-varying dark energy parameters ($w, w_0, w_a$) remain remarkably robust and stable across all redshift exclusion tests. This confirms that the observed cosmological signal, specifically the preference for dynamical dark energy is a global feature uniformly distributed across the HSC Y3 redshift range, rather than an artifact driven by anomalous systematics in any single tomographic bin.

While the dark energy sector remains highly stable, we observe slightly larger shifts in the structure growth parameters ($S_8$ and $\sigma_8$) when excluding the higher-redshift bins ($z_3$ or $z_4$) in the extended cosmological models. As illustrated by the 2D posterior distributions in Appendix \ref{2D posteriors for internal tests} (Figure \ref{fig: wCDM result Appendix} and Figure \ref{fig: w0waCDM result Appendix}), there is a prominent degeneracy between the structure growth amplitude and the photometric redshift mean shifts of the high-redshift bins ($\Delta z_3$ and $\Delta z_4$). Removing the $z_3$ or $z_4$ bin from the data vector effectively removes its associated $\Delta z$ nuisance parameter from the likelihood. Breaking this degeneracy causes a slight shift in the marginalized 1D posteriors for $S_8$ and $\sigma_8$. This known photo-$z$ degeneracy seems however decoupled from the dark energy equation of state, preserving the overall integrity of our primary $(w_0, w_a)$ constraints.

\subsection{Impact of Tomographic Bin Exclusion on Observables}
\label{subsec:bin_exclusion_2pcf}

\begin{figure}[htbp]
\centering
\includegraphics[width=\columnwidth]{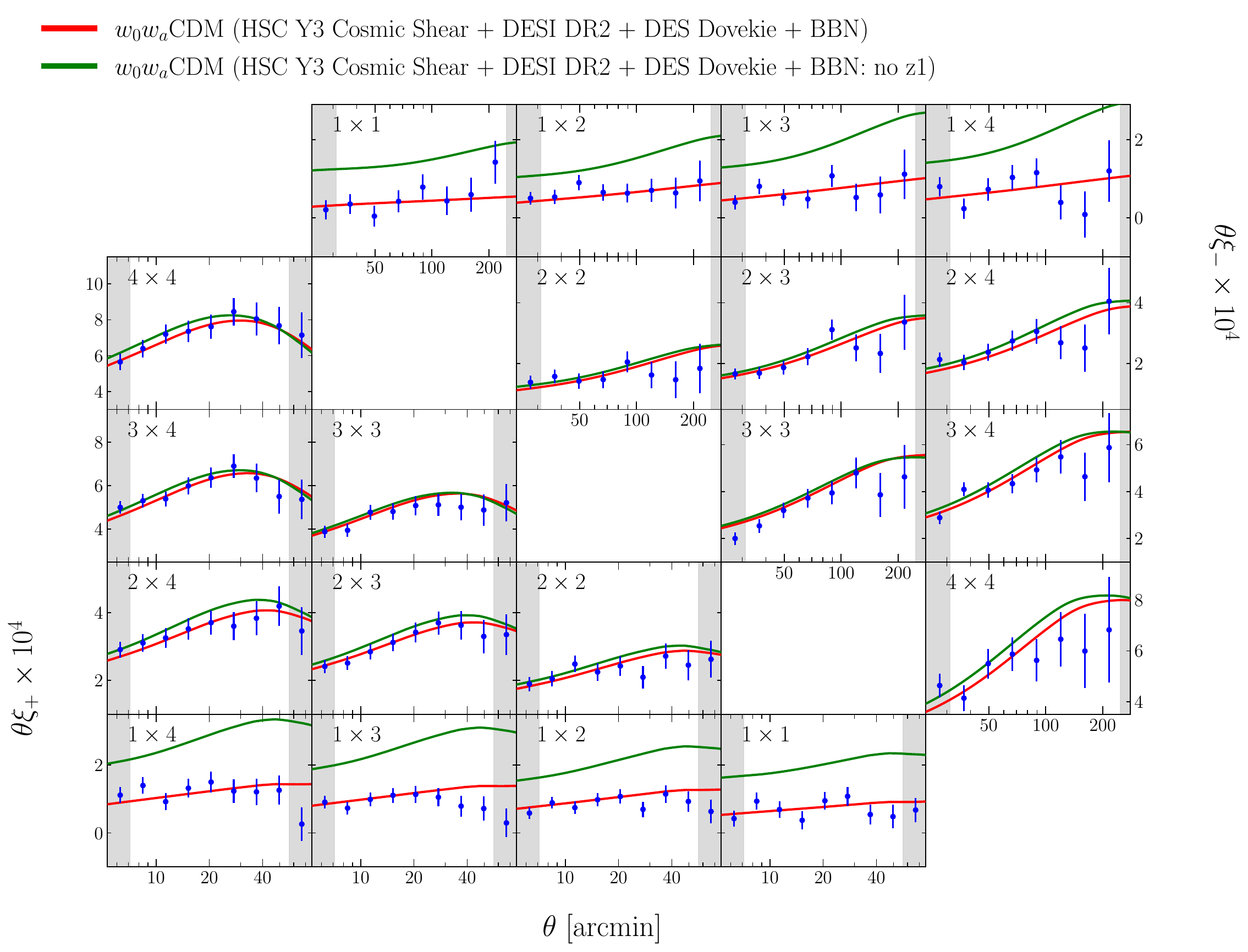}
\caption{The ten 2PCFs including four autocorrelations and six cross-correlations between the four tomographic redshift bins
(labeled with 1–4). This plot shows the 2PCFs on scales $5.3 < \xi_+ < 76$ [arcmin] for $\xi_+$, and $23.2 < \xi_- < 248$ [arcmin] for $\xi_-$. The unshaded region refers to the fiducial scale cut: $7.1 < \xi_+ < 56.6$ [arcmin] for $\xi_+$, and $31.2 < \xi_- < 248$ [arcmin] for $\xi_-$ . The errorbars are estimated with mock catalogs. The solid lines show the model predictions at the maximum-posterior (MAP) sample of each chain of our $w_0w_a$CDM analysis.}
\label{fig: w0waCDM result to verify result for no z1}
\end{figure}

To further investigate the sensitivity of our inference to the lowest redshift interval, we evaluate how the exclusion of specific tomographic bins alters the predicted observables. The Figure \ref{fig: w0waCDM result to verify result for no z1} presents the measured $\xi_+$ correlation functions across all tomographic bins, comparing the data against two distinct theoretical predictions: our fiducial joint $w_0w_a$CDM baseline (red line) and a modified inference run where the first tomographic bin ($z_1$) is entirely excluded (green line). 

When the $z_1$ data is omitted from the likelihood, the resulting best-fit cosmology overpredicts the clustering amplitude across all correlation panels involving the first bin, most notably in the $1\times 1$, $1\times 2$, $1\times 3$, and $1\times 4$ configurations (Figure \ref{fig: w0waCDM result to verify result for no z1}). In contrast, the fiducial model provides an excellent, well-calibrated fit to these low-redshift data points. This visual deviation in the predictive data space highlights the critical role the lowest redshift bin plays in anchoring the overall amplitude of the cosmic shear signal at late times. While the marginalized cosmological parameters remain statistically robust to the removal of $z_1$, this verification plot underscores the raw constraining weight contributed by the $z_1$ cross-correlations in preventing the model from overestimating the low-redshift clustering amplitude.

\subsection{Expansion History and Residuals}
\label{subsec:residuals}

\begin{figure*}[tbp]
\centering
\includegraphics[width=\textwidth]{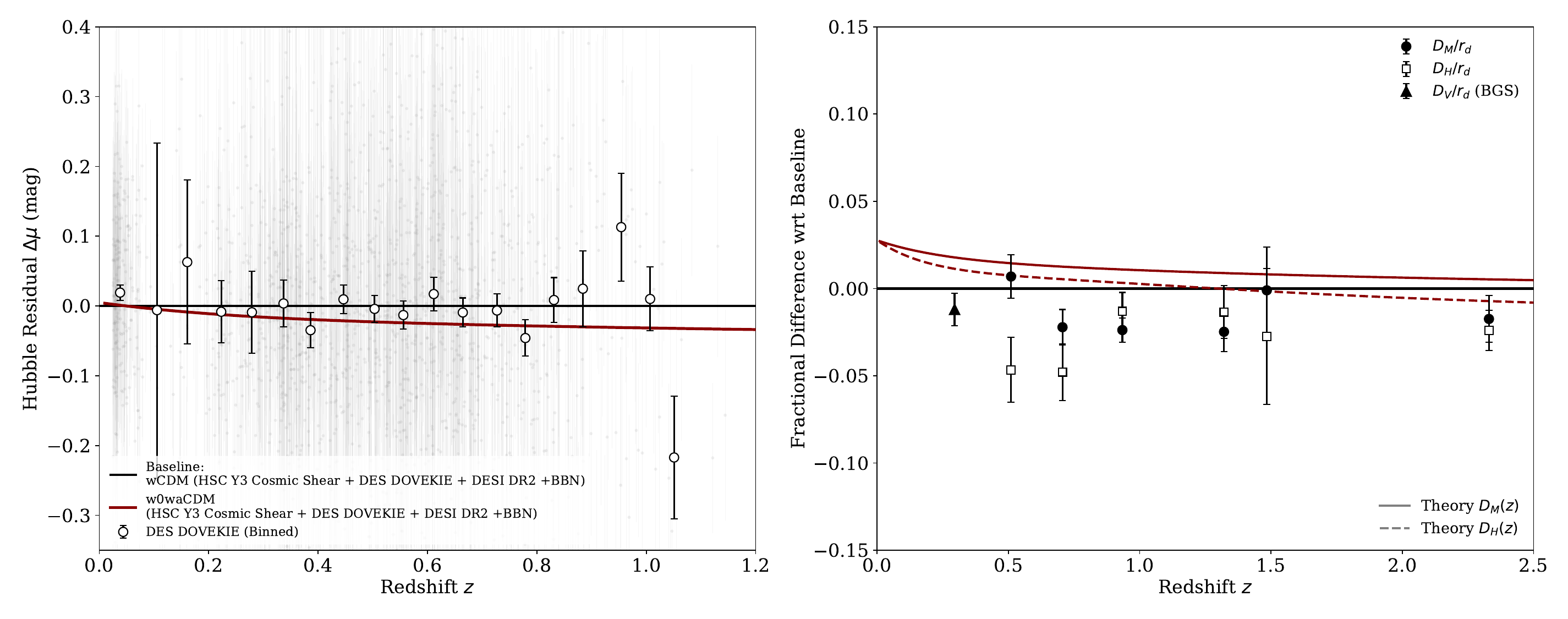}
\caption{To quantify the goodness-of-fit and expansion history preferences for our baseline models, we evaluate the joint fit across the HSC Y3 cosmic shear, DES DOVEKIE supernovae, DESI DR2 BAO, and BBN datasets. While the $w$CDM model provides an excellent overall fit to the combined data (total $\chi^2 = 1799.05$ for $1979$ degrees of freedom), expanding the parameter space to the $w_0w_a$CDM model reveals a $2.4\sigma$ statistical preference for a time-varying dark energy equation of state over a constant $w$. This dynamical behavior is visually reflected in the background distance residuals and is driven primarily by the external geometric distance ladders, corroborating recent findings by the DESI collaboration.}
\label{fig: model_fit_to_SN_BAO_data}
\end{figure*}

Finally, we analyze the preferred background expansion history. The Figure \ref{fig: model_fit_to_SN_BAO_data} displays the Hubble residuals ($\Delta \mu$) and the fractional differences in the comoving ($D_M$) and Hubble ($D_H$) distances relative to a baseline model. 

When HSC-Y3 cosmic shear data is combined with external datasets, the allowed models are heavily anchored to the baseline, forcing the fractional differences in $D_M(z)$ and $D_H(z)$ to remain near zero across the entire redshift range probed by the DES DOVEKIE supernovae (Figure \ref{fig: model_fit_to_SN_BAO_data}). This visually reinforces that the external distance ladders are responsible for breaking the lensing degeneracies, pulling the otherwise weakly-constrained $w(a)$ parameters into the $2.4\sigma$ dynamical configuration while cosmic shear accurately anchors the amplitude of structure growth.

\section{Discussion}\label{Discussion}

\subsection{Limitations of the Current Modeling and Future Improvements}\label{Limitations and Future Work}

The modeling pipeline adopted in this work, comprising the BACCO linear emulator (for $w$CDM) or CAMB (for $w_0w_a$CDM), HMCode2016 for nonlinear clustering and baryonic feedback, and the TATT intrinsic alignment model, was originally designed and validated by \citet{2023PhRvD.108l3518L} for a flat $\Lambda$CDM cosmology. In this work, we extend the analysis to dynamical dark energy scenarios, namely $w$CDM and $w_0w_a$CDM. Although the pipeline can be run for these models, several components were not built or calibrated for them. Our current results should therefore be read as a first step, and we expect that targeted changes to the modeling choices can yield more accurate and tighter constraints on the dark energy equation of state. 

First, the BACCO emulator is trained over a finite cosmological parameter space. Its coverage of the dark energy parameters $w_0$ and $w_a$ is considerably narrower than the broad, uninformative priors typically adopted in the $w_0w_a$CDM analyses. Regions of parameter space outside the emulator's training range must either be excluded through restrictive priors, which can artificially tighten the posteriors, or be extrapolated, which degrades the accuracy of the predicted linear power spectrum. Therefore, for this case, we have replaced the emulator with a direct Einstein-Boltzmann solver such as CAMB \citep{Lewis_2000} or CLASS \citep{2011arXiv1104.2932L}. 

Second, HMCode2016 \citet{Mead_2016} was calibrated primarily against $\Lambda$CDM simulations, and its accuracy for evolving dark energy models, especially for $w_0w_a$CDM cosmologies far from $(w_0, w_a) = (-1, 0)$, needs to be further validated. Its updated version, HMCode2020 \citet{Mead_2021}, was calibrated against simulations that include dynamical dark energy and massive neutrinos. It also describes baryonic feedback with a physically motivated AGN heating temperature parameter ($\log_{10} T_{\mathrm{AGN}}$) rather than the empirical $(A_b, \eta_b)$ relation used here. Simulation-based nonlinear emulators that explicitly span $w_0$ and $w_a$ offer a further alternative, for example EuclidEmulator2 \citep{2021}, the Mira-Titan emulators \citep{Moran_2022}, and the BACCO nonlinear emulator \citep{Angulo_2021} together with its baryonification model \citep{Aric__2021}. Since dark energy affects both the growth of structure and the shape of the nonlinear power spectrum, and since these effects are partly degenerate with baryonic suppression on small scales, a more consistent treatment of nonlinear and baryonic physics is expected to reduce modeling bias. It may also allow the use of smaller angular scales, improving the constraining power.

Third, the scale cuts and the model validation tests on which we rely \citet{2023PhRvD.108l3518L} were defined for $\Lambda$CDM. The level of modeling bias and projection effects in extended cosmologies could potentially differ, since the additional parameters $w_0$ and $w_a$ introduce new degeneracies with $\Omega_{\rm m}$, $S_8$, the intrinsic alignment amplitudes, and the photometric redshift shifts of the high-redshift bins. A dedicated validation using synthetic data vectors generated in $w$CDM and $w_0w_a$CDM cosmologies is therefore required to confirm that the $<0.3\sigma$ bias criterion is still satisfied.

Improving these modeling choices, and quantifying their impact on the inferred dark energy parameters, is the focus of future work.

\section{Conclusion}\label{Conclusions}

In this work, we have extended the baseline $\Lambda$CDM cosmological analysis of the Hyper Suprime-Cam (HSC) Year 3 real-space cosmic shear two-point correlation functions ($\xi_{\pm}$) to constrain extended dark energy models. Motivated by recent hints from DESI BAO and Type Ia Supernovae that dark energy may evolve with time, we conducted a consistency test using weak lensing constraints. By combining the late-time structure growth measurements from HSC with CMB-independent background expansion probes—namely Baryon Acoustic Oscillations from DESI DR2, Type Ia Supernovae from the DES DOVEKIE sample, and Big Bang Nucleosynthesis (BBN) priors, we effectively break the geometric degeneracies inherent in weak gravitational lensing alone. 

\noindent Our primary findings are as follows:

\begin{itemize}
    \item In the context of the $w$CDM model, using HSC Y3 cosmic shear data alone, we observe a strong degeneracy in the $\Omega_{\rm m} - w$ plane, which artificially pulls the structure growth parameter $S_8$ to lower values. However, the inclusion of external distance calibrators tightly constrains the dark energy equation of state to $w = -0.936^{+0.074}_{-0.067}$, consistent with a cosmological constant. Once the geometric degeneracy is broken, the joint constraint yields $S_8 = 0.774^{+0.027}_{-0.034}$, similar to the fiducial HSC-Y3 $\Lambda$CDM value. Our constraints also show excellent agreement with recent independent analyses, such as \citet{2026PhRvD.113j3530Z}.
    
    \item In the context of the $w_0w_a$CDM model and with inclusion of geometric probes, we find $w_0 = -0.832^{+0.071}_{-0.059}$ and $w_a = -1.23^{+0.66}_{-0.44}$, representing a $2.4\sigma$ preference for dynamical dark energy, consistent with recent DESI findings. The structure growth parameter remains robust at $S_8 = 0.772^{+0.031}_{-0.030}$. This indicates that the HSC-Y3 lensing amplitude is fundamentally robust to the assumed dark energy model, and that the dynamical dark energy signal is primarily driven by the geometric probes.
    
    \item We compared our HSC-anchored constraints against the multi-probe DES Y5 ($3\times2$pt + SN + eBOSS) analysis. The marginalized posteriors in both the $w$CDM and $w_0w_a$CDM parameter spaces exhibit statistical consistency, tracing a similar degeneracy direction that favors $w_0 > -1$ and $w_a < 0$.

    \item Through a suite of internal consistency checks, including a leave-one-out tomographic bin analysis, we demonstrated that our cosmological constraints are not driven by isolated systematics in any single redshift interval. Furthermore, predictive verification of the observables confirms that the lowest redshift data acts as a crucial anchor for the clustering amplitude, preventing the over-prediction of the low-redshift $\xi_+$ signal. Finally, the inclusion of extended dark energy parameters does not induce pathological degeneracies with our systematic nuisance parameters, such as intrinsic alignments ($A_I$) or photometric redshift uncertainties ($\Delta z$).
\end{itemize}

Our analysis of the HSC Y3 cosmic shear data demonstrates that while background geometric probes show an intriguing preference for evolving dark energy, the weak lensing amplitude of structure growth remains robust. The ability to decouple late-time clustering from the background expansion history highlights the synergy between the combination of deep, high-resolution imaging surveys with spectroscopic distance measurements. As we transition into the era of Stage-IV surveys, such as Rubin LSST and Euclid, this demonstrated robustness of the lensing amplitude to dynamical dark energy extensions will serve as a foundational consistency test to further tighten the bounds on the dynamic evolution of our Universe.

\begin{acknowledgments}

We thank Masamune Oguri, Sunao Sugiyama, Ryo Terasawa, and Tianqing Zhang for helpful discussions, and in particular Tianqing Zhang for providing the HSC Y3 + SDSS $3\times2$pt chains used in our comparison. We are also grateful to the HSC weak lensing working group for valuable feedback throughout this analysis. ShM thanks Mr. Sourav Das for his invaluable and countless insightful discussions, and constant moral support, as well as colleagues and seniors for their collaborative spirit. We acknowledge the Inter-University Centre for Astronomy and Astrophysics (IUCAA) for providing the high-performance computing facility ‘Pegasus’, which was instrumental in executing the MCMC analyses presented in this work. 
\end{acknowledgments}

\section*{Data Availability}

The Hyper Suprime-Cam Year 3 shear catalog and summary statistics are publicly available at \href{https://hsc-release.mtk.nao.ac.jp/doc/index.php/wly3/}{HSC Data Release Website}. The specific CosmoSIS \texttt{.ini} pipelines, and the PolyChord MCMC chains generated for the extended $w$CDM and $w_0w_a$CDM analyses in this work will be made publicly available via GitHub at \href{https://github.com/CosmicShreya/HSCY3-dynamical-dark-energy-chains}{HSCY3-dynamical-dark-energy-chains} upon the submission of this manuscript.
\bibliographystyle{apsrev4-2}

\bibliography{refs}

\appendix

\section{2D posteriors for internal tests}\label{2D posteriors for internal tests}

\begin{figure*}
\centering
\includegraphics[width=\textwidth]{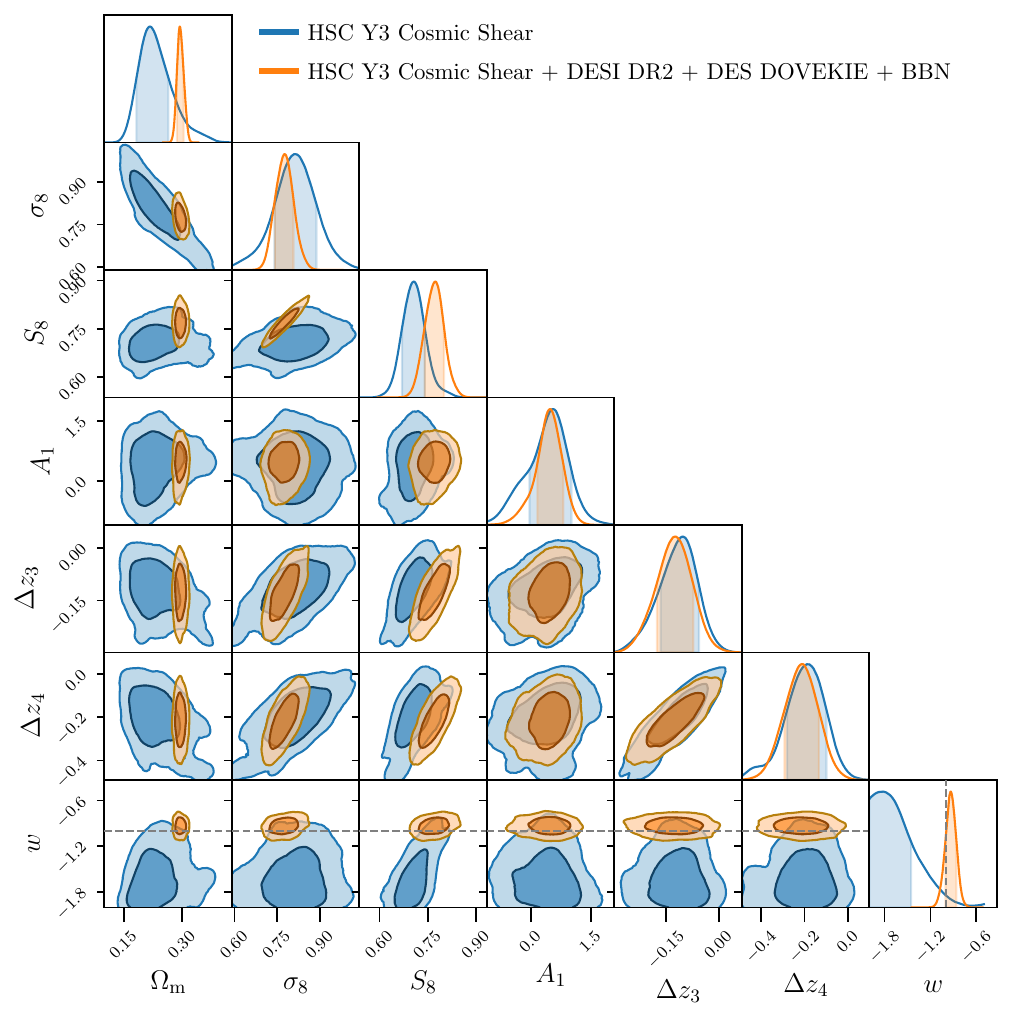}
\caption{Extended 1D and 2D marginalized posterior distributions for the $w$CDM model using Hyper Suprime-Cam Year 3 data. This plot explicitly illustrates the correlations between the primary cosmological parameters ($\Omega_{\rm m}, \sigma_8, S_8$, and the dark energy equation of state $w$) and the primary astrophysical and observational nuisance parameters. Specifically, we display the linear intrinsic alignment amplitude ($A_1$) and the photometric redshift mean shifts across the four tomographic bins ($\Delta z_1, \Delta z_2, \Delta z_3, \Delta z_4$). The different colors represent the different likelihood combinations mentioned in the legend. The stability of the nuisance parameter posteriors upon the inclusion of external geometric probes confirms that our systematics mitigation remains robust and largely independent of the background cosmology.}
\label{fig: wCDM result Appendix}
\end{figure*}

In this appendix, we present the extended marginalized posterior distributions for our primary cosmological models. While the main text focuses on the headline constraints for structure growth and the dark energy equation of state, it is critical to verify that these extended parameter spaces do not introduce pathological degeneracies with our astrophysical and observational nuisance parameters. 

\begin{figure*}
\centering
\includegraphics[width=\textwidth]{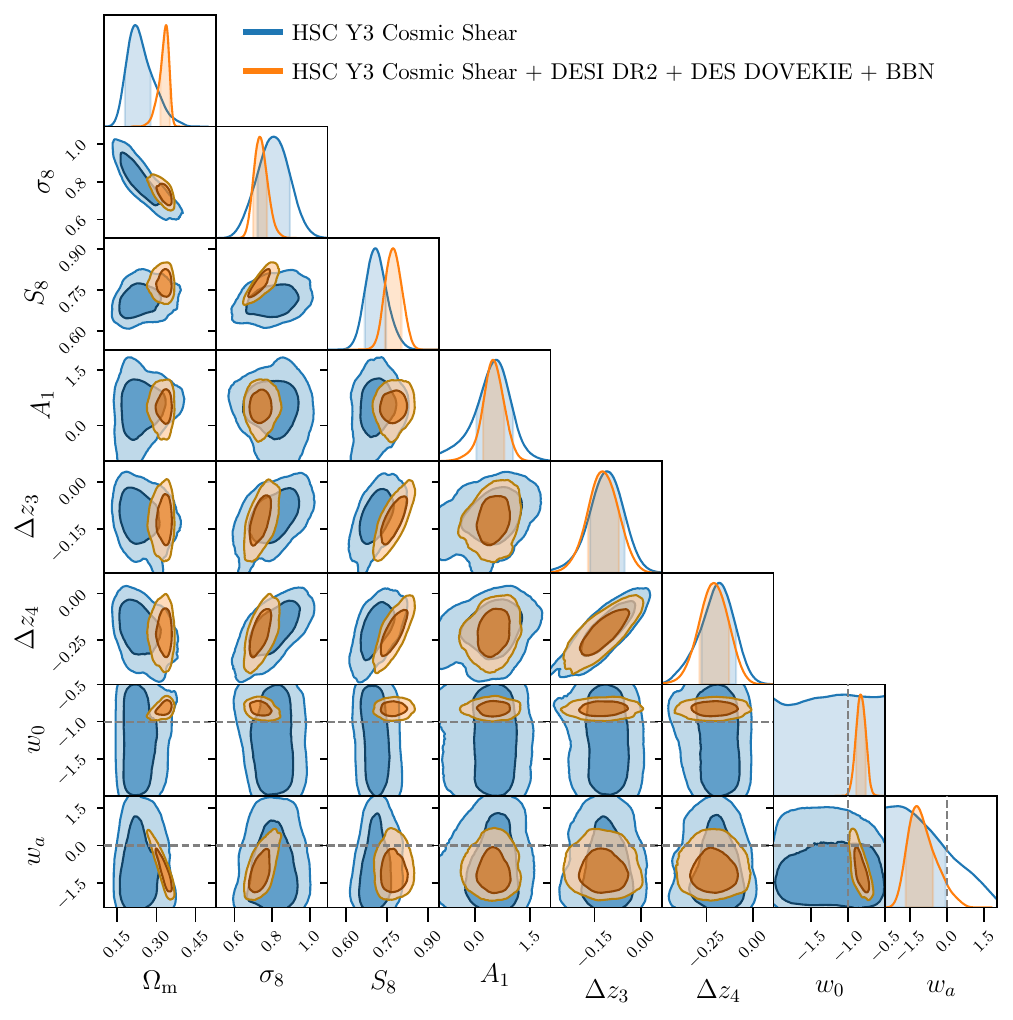}
\caption{Extended 1D and 2D marginalized posterior distributions for the dynamical $w_0w_a$CDM model using Hyper Suprime-Cam Year 3 data. Similar to Figure A1, this plot demonstrates the parameter space encompassing the cosmological parameters ($\Omega_{\rm m}, \sigma_8, S_8, w_0, w_a$), the intrinsic alignment amplitude ($A_1$), and the photometric redshift uncertainties ($\Delta z_1, \Delta z_2, \Delta z_3, \Delta z_4$). The different colors denote the likelihood combinations listed in the legend. Crucially, the time-varying dark energy parameters ($w_0, w_a$) exhibit no pathological degeneracies with the systematic parameters, ensuring that the $2.4\sigma$ preference for dynamical dark energy is not artificially driven by unmodeled observational biases.}
\label{fig: w0waCDM result Appendix}
\end{figure*}

Figure \ref{fig: wCDM result Appendix} displays the extended posteriors for the $w$CDM model, highlighting the correlations between the cosmological sector and the intrinsic alignment amplitude ($A_1$) alongside the photometric redshift mean shifts ($\Delta z_{1-4}$). Figure \ref{fig: w0waCDM result Appendix} presents the corresponding parameter space for the dynamical $w_0w_a$CDM model. Across both models, the nuisance parameters remain stable upon the inclusion of external geometric probes. They exhibit no strong unphysical correlations with the dark energy sector.

\end{document}